\documentclass[aps,prb,10pt,reprint,amsmath,amssymb,amsfonts,showkeys,superscriptaddress]{revtex4-1}
\pdfoutput=1
 
\usepackage{graphicx,graphics}
\usepackage{latexsym,verbatim}
\usepackage{bm}
\usepackage{natbib}
\usepackage[breaklinks=true,colorlinks,citecolor=blue,linkcolor=blue,urlcolor=blue]{hyperref}
\usepackage{siunitx}
\usepackage[utf8]{inputenc}
\usepackage{physics}
\usepackage[percent]{overpic}
\usepackage{tikz}

\begin{document}
\author{Thomas Benjamin Smith}
\affiliation{School of Physics and Astronomy, University of Manchester, Manchester, M13 9PY, United Kingdom}
\author{Iacopo Torre}
\affiliation{ICFO-Institut de Ci\`{e}ncies Fot\`{o}niques, The Barcelona Institute of Science and Technology, Av. Carl Friedrich Gauss 3, 08860 Castelldefels (Barcelona),~Spain}
\author{Alessandro Principi}
\affiliation{School of Physics and Astronomy, University of Manchester, Manchester, M13 9PY, United Kingdom}
\title{Edge modes and Fabry-Perot Plasmonic Resonances in anomalous-Hall Thin Films}

\begin{abstract}
We study plasmon propagation on a metallic two-dimensional surface partially coated with a thin film of anomalous-Hall material. The resulting three regions, separated by two sharp interfaces, are characterised by different 
Hall conductivities but identical 
normal conductivities.  
A single bound mode is found, which can localise to either interface and has an asymmetric potential profile across the region. 
For propagating modes, we calculate the reflection and transmission coefficients through the magnetic region. 
We find Airy transmission patterns with sharp maxima and minima as a function of the plasmon incidence angle. 
The system therefore behaves as a high-quality filter.
\end{abstract}


\maketitle

\section{Introduction}

Much theoretical~\cite{Huang:2017,Wunsch_njp_2006,Polini_prb_2008,Grigorenko_nature_phot_2012,Yan_nature_phot_2013,Principi_prb_2013,Principi_prb_2013b,Principi_prb_2014}  and experimental~\cite{Fei_nature_2012,Chen_nature_2012,Woessner:2014,Gonzalez_nature_nano_2016,Ni_nature_phot_2016,Lundeberg_nature_mater_2016,Bezares_nanolett_2017,Low_prb_2018,Dias_prb_2018,Ni_nature_mater_2018,Ni_nature_2018,
Alcaraz_science_2018,basov:science:2016} work of recent years has highlighted the potential of graphene's two-dimensional (2D) surface plasmon polaritons in next-generation transistors, emitters and detectors. Their extraordinary properties include, but are not limited to, small confinement scales at high-field,\cite{Gonzalez_nature_nano_2016,Alcaraz_science_2018} long lifetimes and low losses,\cite{Woessner:2014,Ni_nature_2018} and gate-tunability of the propagation wavelength.\cite{Grigorenko_nature_phot_2012,Koppens:2011}

Plasmons are high-frequency, electronic density waves that occur at frequencies at which the metal dielectric function vanishes. The long lifetimes of plasmons at small momenta stem from their inability, without the aid of impurities and phonons,\cite{Giuliani:2005,Principi_prb_2013,Principi_prb_2013b} to excite single-electron-hole pairs. On the other hand, their small confinement scales are due to their weak self-interaction, which suppresses any incoherence-causing diffraction.\cite{Oulton_iop_2008} Surface plasmons are exponentially localised to interfaces between a metal and a dielectric (or the vacuum).

Provided that the surface in question is capable of hosting metallic conducting electronic states, plasmonic oscillations may be supported. Such systems include the 2D surface of a general 3D topological insulator.\cite{Qi:2011,Hasan:2010,Salehi:2015,Yu:2010,Wang:2015,Zhang:2017,Dybko_prb_2017,Hasan:2013,Yi:2014,Qi_rmp_2011}
In such cases, the low-energy electronic states possess linear dispersions and behave as massless Dirac fermions.
When time reversal symmetry is broken by, {\it e.g.}, a local magnetisation~\cite{Yu:2010,Chang:am:2013} or a magnetic field\cite{Cheng_prl_2010,Chang:am:2013,Qiao:2014,Zhang:2015} then gaps open in the surface band structure. As a result, when the Fermi energy is tuned to reside in such a magnetisation gap, electrons are characterised by a finite, frequency independent Hall conductance, in units of $e^2/h$,\cite{Chang:am:2013,Liu_prl_2013} where $e$ and $h$ are the electronic charge and Planck's constant respectively. The Hall conductivity decreases, and eventually vanishes, when the Fermi energy is pushed far away from the middle of the magnetisation gap, in either the conduction or valence band. Regardless, all of these situations have the effect of causing the emergence of a frequency independent Hall conductance of the order of $e^2/h$. 
A similar phenomenon occurs in spin-orbit coupled metallic thin films in the presence of a finite magnetisation or magnetic disorder.\cite{Nagaosa_rmp_2010}

\begin{figure}
\centering
\includegraphics[trim=5cm 19cm 1.5cm 4.5cm,width=\linewidth]{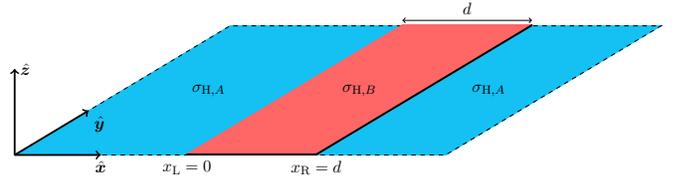}
\caption{(Colour on-line)
A diagram of the system under consideration. The central region $B$ of width $d$ is the thin 2D magnetic film. It is bounded on both sides by sharp interfaces and is characterised by a Hall conductance of $\sigma_{{\rm H},B}$, whilst in the other two regions the Hall conductance is $\sigma_{{\rm H},A}$. The $\hat{\bm{y}}$ direction is assumed to extend uniformly to $\pm\infty$ and the $\hat{\bm{x}}$ direction to $-\infty$ for $x<0$ and $+\infty$ for $x>d$ with the thin-film located at $0\leq x\leq d$. (Note that, due to the mirror symmetry of the problem in $x$, this configuration is identical to any other so long as $x_R-x_L=d$.) Finally, the dielectric environment is assumed uniform and equivalent to air such that $\epsilon_1=\epsilon_2=1$.}
\label{fig:diagram}
\end{figure}

In this paper, we investigate a ``2D thin-film geometry'', whereby a narrow region of a 2D metallic surface exhibits a finite Hall conductivity (due to, {\it e.g.} a local nonvanishing magnetisation)--see Fig.~\ref{fig:diagram}.
The interfaces separating the three regions are assumed to be sharp relative to the plasmon wavelength so that boundary effects may be ignored. Furthermore, the conductivity is assumed to be local so that it does not depend on the plasmon wavevector. Finally, it is assumed to be isotropic so that the conductivity tensor may be decomposed into a normal diagonal part and an antisymmetric, off-diagonal Hall part. By allowing for these approximations, we are implicitly assuming that the inverse of the plasmon wavevector is much larger than both the magnetic thin-film size and the domain-wall length.
(In passing, we note that the impact of sharp variations of the (valley-)Hall conductivity, at domain walls between AB and BA regions, on the plasmons of bilayer graphene has been studied in Ref.~\onlinecite{Hasdeo:2017}.)

The frequency-independent off-diagonal Hall conductivity therefore varies step-wise between the three regions. Conversely, we assume that the normal frequency-dependent conductivity is spatially independent, {\it i.e.} it assumes the same value across the magnetised thin film. This approximation is justified by the fact that the normal conductivity is much less sensitive to variations of the magnetisation whereas the Hall conductivity, under the same conditions, jumps from zero to a finite value. 
Although simplified, this model captures the fundamental physics of the problem, and lends well to experimental testing. There, the typical plasmon wavelength is of the order of $2\pi/q_{\rm p} \sim 100~{\rm nm}$.\cite{Woessner:2014}

The paper is organised into four parts. Firstly, a semi-classical model for plasmon propagation is derived. 

Secondly, the dispersions, lifetimes and potential profiles of bound interface states, which exist within the thin-film region and are exponentially localised at the interfaces, are found.
Plots of these quantities with varying interface separation and wavevector parallel to the interfaces are then shown and discussed.

Thirdly, de-localised states propagating through the thin film are investigated.
Their frequency is given by the classical 2D plasmon frequency $\omega_{\rm p}(q_{\rm p})$ and are assumed to be undamped.
The reflection and transmission coefficients that characterise the region are found and are used to plot transmittance spectra for varying region thickness and Hall conductance of the thin-film.

Finally, typical experimental conditions along with any potential applications and possible extensions are discussed.

\section{The Semi-classical Model}
To highlight the fundamental physics at play, and without the pretense of describing a particular experimental realisation of the setup, we consider the simplest possible model: a conducting 2D surface of helical massless Dirac fermions. In addition, the dielectric environment is assumed to be air. In the magnetised strip, $0<x<d$, Dirac fermions acquire a finite mass and the Hall conductivity becomes non-zero. The system is assumed isotropic in the ${\hat {\bm y}}$-direction. Such a description applies, e.g., to electrons at the surface of a thick 3D topological insulator~\cite{Qi:2011} (such that its surfaces are electrostatically decoupled), once the correct dielectric environment is taken into account. It can also qualitatively describe plasmons in spin-orbit coupled metallic thin films.\cite{Nagaosa_rmp_2010}

Surface-state electrons are described with a continuum massless-Dirac-fermion model. We assume the system to be $n$-doped with a surface carrier density $n$, which defines a Fermi wavevector and energy $k_{\rm F} = \sqrt{4\pi n/N_{\rm f}}$ and $\varepsilon_{\rm F} = \hbar v_{\rm F} k_{\rm F}$, respectively. Here $N_{\rm f}$ is the number of fermion flavors and $v_{\rm F}$ is the density-independent Fermi velocity.

We employ a continuum semi-classical description, whereby the electronic flow is modelled by collective properties, {\it i.e} the deviation of the charge density from its equilibrium value $\rho(\bm{r},t)$ and the charge current $\bm{j}(\bm{r},t)$. The two are connected by the continuity equation:
\begin{equation} \label{eqn:conteqn}
-i\tilde{\omega} \rho(\bm{r},\tilde{\omega})+\bm{\nabla}\cdot\bm{j}(\bm{r},\tilde{\omega})=0,
\end{equation}
whereas the response of the charge current to the (self-induced) electric field obeys the linear-response Ohm's law:
\begin{equation} \label{eqn:Euleqn}
j_i(\bm{r},\tilde{\omega}) = \sigma_{ij}(\bm{r},\tilde{\omega}) E_j(\bm{r},z=0,\tilde{\omega}).
\end{equation}
Here Einstein's summation convention for Roman indices (standing for in-plane Cartesian components) is understood, and the local conductivity $\sigma_{ij}(\bm{r},\tilde{\omega})$ is determined microscopically. 
In Eqs.~(\ref{eqn:conteqn}) and~(\ref{eqn:Euleqn}), $\rho(\bm{r},\tilde{\omega})$, $j_i(\bm{r},\tilde{\omega})$, $\sigma_{ij}(\bm{r},\tilde{\omega})$ and $E_j(\bm{r},z=0,\tilde{\omega})$ are the Fourier components in complex frequency space, $\tilde{\omega}=\omega+i/\tau$ with $\tau=1/\Gamma$, of the charge density, current, conductivity and electric field, respectively, and $\bm{r}=x\hat{\bm{x}}+y\hat{\bm{y}}$.

We assume the system to be locally isotropic. Therefore:
\begin{equation} \label{eqn:sigma}
\sigma_{ij}(\bm{r},\tilde{\omega})\equiv \sigma(\tilde{\omega})\delta_{ij}+\sigma_{\rm H}(x,\tilde{\omega})\varepsilon_{ij}.
\end{equation}
where $\delta_{ij}$ and $\varepsilon_{ij}$ are the Kronecker-delta and 2D antisymmetric Levi-Civita symbols, respectively.
To simplify the further analysis, we assume $\sigma(x)$ to be spatially independent, whereas the Hall component $\sigma_{\rm H}(x)$ varies stepwise across the interfaces:
\begin{equation} \label{eqn:sigmaH}
\sigma_{\rm H}(x,\tilde{\omega})=\sigma_{{\rm H},A}[\theta(x-d)+\theta(-x)]+\sigma_{{\rm H},B}\theta(d-x)\theta(x).
\end{equation}
Hereafter we assume $\sigma_{{\rm H},A}$ and $\sigma_{{\rm H},B}$ to be frequency independent and the interfaces to be infinitely sharp. This approximation is valid as long as the typical wavelengths of the problem are much longer than the length scales of the interface.
For the sake of definitiveness, we write:
\begin{equation}
\sigma(\tilde{\omega})=\frac{i\mathcal{D}}{\tilde{\omega}+i\gamma},
\end{equation}
where ${\cal D}$ is the Drude weight and $\gamma=1/\tau_{\rm sc}$ is the scattering rate of the underlying electronic carriers.

The problem as defined by the constitutive relations~(\ref{eqn:conteqn})-(\ref{eqn:sigmaH}) is solved together with the self-induced 3D Poisson's equation:
\begin{equation}\label{eqn:Poisseqn}
\nabla^2\phi(\bm{r},z,\tilde{\omega})=-4\pi \rho(\bm{r},\tilde{\omega})\delta(z). 
\end{equation}
Note that, whilst electrons are bounded to the 2D surface, the electric potential extends to the whole 3D space. To determine the plasmons of the heterostructure, we assume that no external electric field is applied, and that ${\bm E}({\bm r},z,t) = -{\bm \nabla}\phi({\bm r},z,t)$.
Since the system is assumed to be translationally invariant in the ${\hat {\bm y}}$-direction, all quantities may be expanded in Fourier components along $\hat{\bm{y}}$, e.g.: $\rho(\bm{r},\tilde{\omega})=\rho(x,\tilde{\omega})e^{iq_yy}$. Thus the problem reduces to that of a 1D well/barrier.

Solving the Poisson's equation given by Eq.(\ref{eqn:Poisseqn}) and taking the Fourier transform of the solution we achieve:\cite{Fetter:1985}
\begin{equation} \label{eqn:FTK0}
\hat{\phi}(q_x)=L(q)\hat{\rho}(q_x),
\end{equation}
where $q=[q_x^2+q_y^2]^{1/2}$ and:
\begin{equation}
L(q)=\frac{4\pi}{(\epsilon_1+\epsilon_2)q}=\frac{2\pi}{q},
\end{equation} 
since the dielectric environment is assumed to be air so that $\epsilon_1=\epsilon_2=1$.  It must be noted here, however, that if the system were a Bismuth based topological insulator this assumption would not hold: the dielectric constants can be orders of magnitude greater than one.\cite{Madelungu:1998}

Fourier-transforming Eq.~(\ref{eqn:conteqn}), and using Eqs.~(\ref{eqn:Euleqn}),~(\ref{eqn:sigma}) and~(\ref{eqn:sigmaH}), we achieve:
\begin{equation} \label{eq:rho_FT}
\hat{\rho}(q_x)=\frac{-i\sigma(\tilde{\omega})}{\tilde{\omega}}q^2\hat{\phi}(q_x)-\frac{\delta\sigma_Hq_y}{\tilde{\omega}}[\phi(0)-\phi(d)e^{-iq_xd}],
\end{equation}
where:
\begin{equation}\label{eqn:conductivites}
\delta\sigma_H=\sigma_{HB}-\sigma_{HA}=\nu\frac{e^2}{2\pi\hbar},
\end{equation}
with $\nu$ as a parameter that characterises the difference between the Hall conductivities of each region. Note that $\nu$ can take either positive or negative values depending on whether the Hall conductivity varies as a well ($\nu<0$) or as a barrier ($\nu>0$).

Combining Eqs.~(\ref{eqn:FTK0}) and~(\ref{eq:rho_FT}), gives a self-consistent relation for the potential in momentum space:
\begin{equation} \label{eqn:mompot}
\epsilon(q,\tilde{\omega})\hat{\phi}(q_x)=-\frac{\delta\sigma_Hq_y}{\tilde{\omega}}L(q)[\phi(0)-\phi(d)e^{-iq_xd}],
\end{equation}
where:
\begin{equation} \label{eq:dielectric_function_bulk}
\epsilon(q,\tilde{\omega})=1+\frac{i\sigma(\tilde{\omega})}{\tilde{\omega}} q^2L(q)=1-\frac{q}{Q(\tilde{\omega})},
\end{equation}
is the dielectric function of the homogeneous surface in the absence of the magnetised strip. $\epsilon(q,\tilde{\omega})$ clearly has a zero at:
\begin{equation}
q = Q(\tilde{\omega}) \equiv \frac{\tilde{\omega}}{2\pi{\cal D}}(\tilde{\omega}+i\gamma)=\frac{\omega}{2\pi{\cal D}}[\omega+i(\gamma+2\Gamma)],
\end{equation}
where $\tilde{\omega}=\omega+i\Gamma$ has been used and any quadratic decay terms are assumed negligible since $\omega\gg\gamma,\Gamma$. We denote, from this point on, the real part of $Q(\omega)$ with $q_{\rm p}(\omega)=\omega^2/(2\pi{\cal D})$, which is the bulk plasmon wavevector. Bound interface plasmons exist for frequencies such that $q_y>q_{\rm p}(\omega)$ whilst propagating solutions for $q_y\leq q_{\rm p}(\omega)$, which we hereafter name the ``continuum region''.

\section{Bounded Interface States}
We start by determining the dispersion and field distribution of interface-localised plasmons, which occur for wavevectors $q_y>q_{\rm p}(\omega)$, {\it i.e.} to the right and below the bulk plasmon continuum. These modes are exponentially localised around the interfaces in a region of size $\xi^{-1}=[q_y^2 - q_{\rm p}^2(\omega)]^{-1/2}$. At these frequencies, $\epsilon(q,\tilde{\omega})\neq0$. We are thus free to divide Eq.~(\ref{eqn:mompot}) by the dielectric function and inverse Fourier transform it back into coordinate space.
We therefore obtain:
\begin{equation}\label{eqn:boundpot}
\phi(x)=-\frac{\delta\sigma_{\rm H}q_y}{\tilde{\omega}}[I(x,q_y,\tilde{\omega})\phi(0)-I(x-d,q_y,\tilde{\omega})\phi(d)],
\end{equation} 
where:
\begin{equation}\label{eqn:Iint}
I(x,q_y,\tilde{\omega})=\mathcal{P}\int_{-\infty}^{+\infty}\frac{dq_x}{2\pi}\frac{L(q)}{\epsilon(q,\tilde{\omega})}e^{iq_xx}.
\end{equation}

To determine the bound states we impose that the potential is continuous at the interfaces. By evaluating Eq.~(\ref{eqn:boundpot}) at $x=0$ and $x=d$ we thus arrive at the following matrix equation:
\begin{equation}\label{eqn:mateqn}
\begin{pmatrix}
1+KI(0,q_y,\tilde{\omega}) && -KI(d,q_y,\tilde{\omega})\\
KI(d,q_y,\tilde{\omega}) && 1-KI(0,q_y,\tilde{\omega})
\end{pmatrix}
\begin{pmatrix}
\phi(0)
\\
\phi(d)
\end{pmatrix}
=0,
\end{equation}
where:
\begin{equation}
K=\frac{\delta\sigma_{\rm H}q_y}{\tilde{\omega}}.
\end{equation}
Note that we have used $I(d,q_y,\tilde{\omega})=I(-d,q_y,\tilde{\omega})$; a relation to be shown subsequently.

Non-trivial solutions of the matrix equation~(\ref{eqn:mateqn}) are found whenever its determinant is zero. This yields the following transcendental equation:
\begin{equation} \label{eqn:dispersion}
\tilde{\omega}^2-\delta\sigma_{\rm H}^2q_y^2[I^2(0,q_y,\tilde{\omega})-I^2(d,q_y,\tilde{\omega})]=0,
\end{equation}
which, ultimately, may only be solved numerically in order to determine the plasmon dispersion relation.

Due to the complex-valued nature of both the frequency and the conductivity, the integral in Eq.~(\ref{eqn:Iint}) may be decomposed into real and imaginary parts, based on the assumption that any term quadratic in the scattering rates $\gamma$ and $\Gamma$ vanishes since $\omega\gg\gamma,\Gamma$, as:
\begin{equation}
I(x,q_y,\omega,\Gamma)={\cal I}(x,q_y,\omega)-i\omega(\gamma+2\Gamma){\cal J}(x,q_y,\omega).
\end{equation}
Then, using this decomposition with $\tilde{\omega}=\omega+i\Gamma$ within Eq.~(\ref{eqn:dispersion}), two simultaneous equations for $\omega$ and $\Gamma$ may be found. The numerical solution of the first yields $\omega$ with which the second may be solved (also numerically) for $\Gamma$, as shall be seen.

Furthermore, the ${\cal I}$ and ${\cal J}$ integrals may be evaluated using contour integration. Due to the the oscillatory Fourier exponential in the integrals, the contours are closed in the upper (lower) half of the complex plane when $x>0$ ($x<0$). In doing so, it becomes apparent that when $q_y<q_{\rm p}(\omega)$ there exist poles at $q_x=\pm iq_0$, where $q_0^2(\omega)=q_y^2-q_{\rm p}^2(\omega)\equiv\xi^{-2}$, and branch cuts along the imaginary axis from $\pm i q_y$ to $\pm i \infty$. On the other hand, when $q_y>q_{\rm p}(\omega)$ the poles exist upon the real axis at $q_x=\pm k_0$, where $k_0^2(\omega)=-q_0^2(\omega)=q_{\rm p}^2(\omega)-q_y^2$, whilst the branch cuts remain unchanged.

Once this is all taken into account we find:
\begin{equation}
\omega^2-\delta\sigma_{\rm H}^2q_y^2\left[{\cal I}^2(0,q_y,\omega)-{\cal I}^2(d,q_y,\omega)\right]=0,
\end{equation}
\begin{multline}
\Gamma+\delta\sigma_{\rm H}^2q_y^2(\gamma+2\Gamma)[{\cal I}(0,q_y,\omega){\cal J}(0,q_y,\omega)\\-{\cal I}(d,q_y,\omega){\cal J}(d,q_y,\omega)]=0,
\end{multline}
where:
\begin{align} \label{eq:I_final}
\{{\cal I}/{\cal J}\}(x,q_y,\omega)=\{{\cal I}/{\cal J}\}_P(x,q_y,\omega)+\{{\cal I}/{\cal J}\}_B(x,q_y,\omega),
\end{align}
with the first terms as the contributions from the poles, which are given by:
\begin{align}
{\cal I}_P(x,q_y,\omega)&=
\begin{cases}
{\displaystyle -2\pi\frac{q_{\rm p}(\omega)}{q_0(\omega)}e^{-q_0(\omega)|x|} },\quad{\rm for}~|q_y|>q_{\rm p}(\omega),
\vspace{0.2cm}\\
{\displaystyle 2\pi\frac{q_{\rm p}(\omega)}{k_0(\omega)}\sin[k_0(\omega)|x|] },~{\rm for}~|q_y|\leq q_{\rm p}(\omega),
\end{cases}
\\
{\cal J}_P(x,q_y,\omega)&=
{\displaystyle 2\pi\left(\frac{q_y^2}{q_0^3(\omega)}+\frac{q_{\rm p}^2(\omega)|x|}{q_0^2(\omega)}\right)e^{-q_0(\omega)|x|} },
\end{align}
and the second terms as the contributions from the branch cuts, which are given by:
\begin{align} \label{eq:I_B_def}
{\cal I}_B(x,q_y,\omega)&=2q_{\rm p}^2(\omega)\int_{0}^{\infty}d\eta\frac{e^{-|q_yx|\cosh(\eta)}}{q_{\rm p}^2(\omega)+q_y^2\sinh^2(\eta)},
\\
{\cal J}_B(x,q_y,\omega)&=4q_{\rm p}(\omega)q_y^2\int_0^\infty d\eta\frac{e^{-|q_yx|\cosh(\eta)}\sinh^2(\eta)}{[q_{\rm p}^2(\omega)+q_y^2\sinh^2(\eta)]^2}.
\end{align}

Note that Eqs.~(\ref{eq:I_final})-(\ref{eq:I_B_def}) depend on the absolute value of $x$. This is due to the fact that, for $x<0$, one closes the contour in the lower half of the complex plane and finds an identical result. This therefore proves the relation $I(x,q_y,\tilde{\omega})=I(-x,q_y,\tilde{\omega})$, which was used earlier.

From Eq.~(\ref{eq:I_final}) it can be seen that the contribution of the poles is exponentially localised around $x$ with characteristic length $\xi$. On the other hand, the branch cut contributions are superpositions of evanescent waves caused by the non-locality of the Coulomb interaction that acts as a transient decay around the interfaces. As such, the behaviour of the plasmonic field is extremely complicated in the immediate vicinity of the interface.

For $q_y\to \infty$ equation (\ref{eqn:dispersion}) simplifies and can be solved analytically yielding:
\begin{equation}
\omega_\infty=\frac{{\cal D}}{|\delta\sigma_{\rm H}|},
\end{equation}
whilst in the limit of small $q_y$, the plasmon tends to the continuum region and so $q_y\to q_{\rm p}$.

The normalised eigenvectors of Eq.~(\ref{eqn:mateqn}), corresponding to the positive frequencies (negative frequencies are simply plasmons moving backwards in time) as obtained from Eq.~(\ref{eqn:dispersion}), are:
\begin{align} \label{eqn:eigenvector}
\begin{pmatrix}
\phi(0)
\\
\phi(d)
\end{pmatrix}
= 
\mathcal{N}_c \begin{pmatrix}
1-KI(0,q_y,\omega)
\\
-KI(d,q_y,\omega)
\end{pmatrix}
,
\end{align}
where $\mathcal{N}_c=\left[(1-KI(0,q_y,\omega,\Gamma))^2+K^2I^2(d,q_y,\omega,\Gamma)\right]^{-\frac{1}{2}}$.

Using Eq.~(\ref{eqn:eigenvector}) in Eq.~(\ref{eqn:boundpot}) allows for solution of the spatially dependent potential relation as:
\begin{equation}
\phi(x)=\frac{[KI(0)-1]I(x)-KI(d)I(x-d)}{\sqrt{\left[I(0)-K^{-1}\right]^2+I^2(d)}},
\end{equation}
where the $q_y$, $\omega$ and $\Gamma$ dependencies of $I(x,q_y,\omega,\Gamma)$ have been dropped for brevity. Note that $K=\delta\sigma_{\rm H}q_y/(\omega+i\Gamma)$.

Although not immediately obvious, this shows that, depending on the combined sign of $\delta\sigma_{\rm H}q_y$, the plasmon will not only prefer to localise to a specific interface but also have . It is in fact the $K$ terms in the numerator that cause this behaviour since $I(x)$ depends on $|q_y|$ and $q_y^2$ only.

In the figures and results to come, we adimensionalise the variables using the linear Dirac dispersion of the electronic system $\varepsilon_{\rm F}=\hbar v_{\rm F}k_{\rm F}$ where $k_{\rm F}=\sqrt{4\pi n}$ and $m v_{\rm F}=\hbar k_{\rm F}$. Note that we set the number of fermion flavors $N_F=1$, which is equivalent to measuring $k_{\rm F}$ in units of $\sqrt{N_F}$. In addition, $\alpha_*=e^2/(\hbar v_F)$ (in CGS units) is the fine-structure constant of said Dirac system. In these units the factor $2\pi\mathcal{D}$ may be expressed as $2\pi\mathcal{D}=\alpha_*\varepsilon_{\rm F}^2/(2\hbar^2k_{\rm F})$ and so:
\begin{equation}
Q(\tilde{\omega})=\frac{Q(\omega+i\Gamma)}{k_{\rm F}}=\frac{2\hbar^2}{\alpha_*\varepsilon_{\rm F}^2}[\omega^2+i\omega(\gamma+2\Gamma)],
\end{equation}

Furthermore, the change in Hall conductivity may be expressed as: $\delta\sigma_{\rm H}=\nu\alpha_*\varepsilon_{\rm F}/(2\pi\hbar k_{\rm F})$, in these units. We take $\alpha_*\approx4$ as a typical value for a 2D Dirac surface state of a 3D-TI.\cite{Zhang:2009,Zhang:2012} On the other hand, $\alpha_*\approx2$ in graphene.\cite{CastroNeto:2009}

\begin{figure}
\begin{center}
\begin{overpic}[width=1.0\columnwidth]{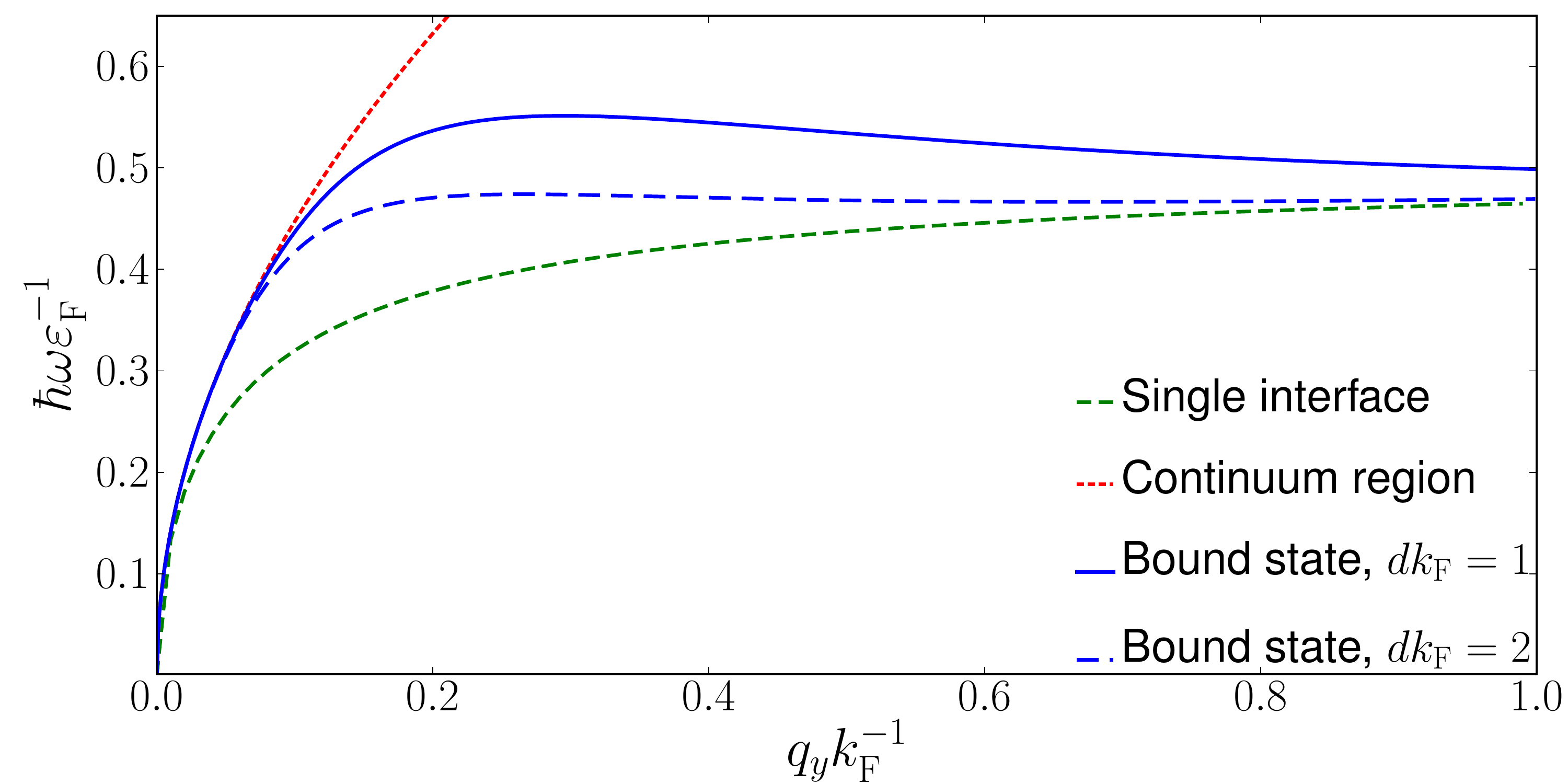}\put(11,45){(a)}
\end{overpic}
\begin{overpic}[width=1.0\columnwidth]{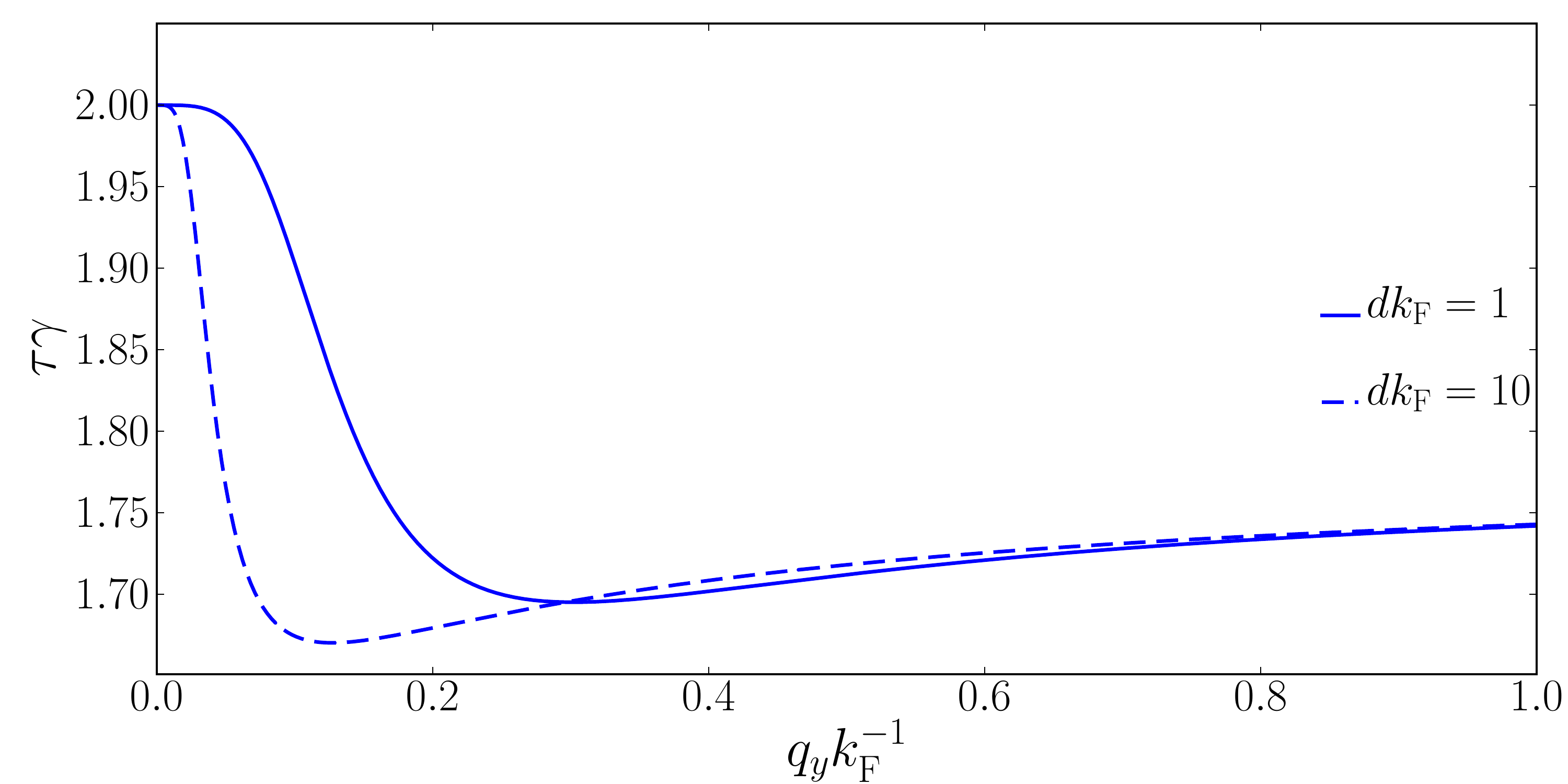}\put(11,45){(b)}
\end{overpic}
\begin{overpic}[width=1.0\columnwidth]{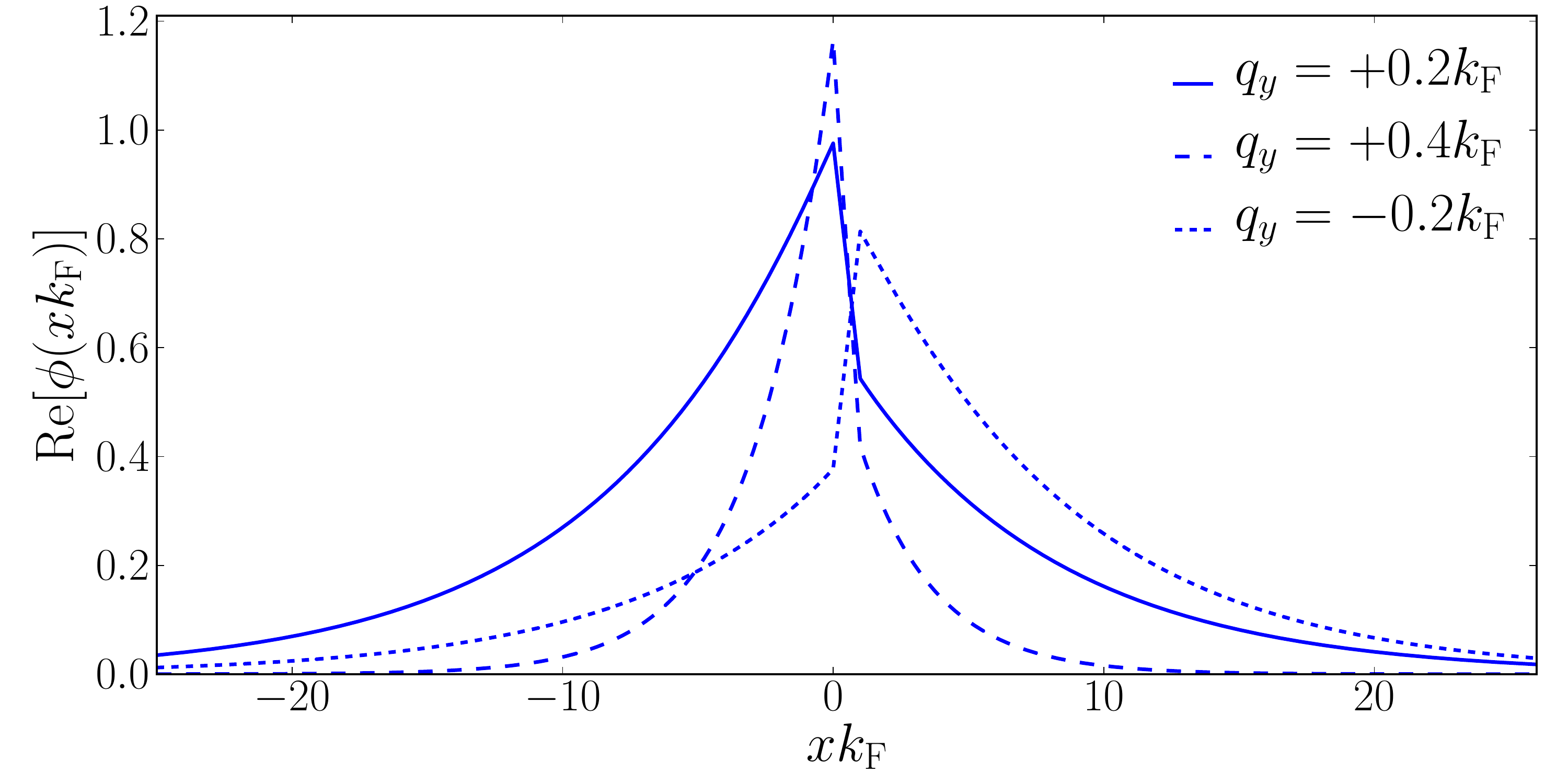}\put(11,45){(c)}
\end{overpic}
\end{center}
\caption{(Colour on-line)
Panel (a): The peak of the energy dispersion of bound states decreases with increasing the interface separation from $dk_{\rm F}=1$ (solid line) to $dk_{\rm F}=2$ (long-dashed line). Here $|\nu|=2$, whereas the short-dashed and dotted lines stand for the single-interface result and the continuum, respectively. Panel (b): Plotted are the plasmon lifetimes as a function of $q_y$ for $dk_{\rm F}=1$ (solid line) and $dk_{\rm F}=10$ (dashed line). The lifetime of the bound state is seen to be greatest at $q_y=0$, have a turning point in $q_y$ that recedes in $q_y$ as $d$ increases and to be roughly constant for all $q_y$ around $\tau\gamma\sim1.8$. Panel (c): The potential of the bound state localizes at one of the two interfaces, depending on the combined sign of $\nu q_y$, with a localization length that is inversely proportional to $q_y$. The solid, long dashed, and small dashed lines correspond to $q_y=+0.2k_{\rm F},+0.4k_{\rm F},-0.2k_{\rm F}$ respectively.}
\label{figBS}
\end{figure}

Fig.~\ref{figBS}(a) shows the energy dispersion $\hbar \omega$ in units of $\varepsilon_{\rm F}$, calculated by numerically solving Eq.~\eqref{eqn:dispersion}, as a function of the momentum $q_y$ along the interface (in units of $k_{\rm F}$). Note that Eq.~\eqref{eqn:dispersion} has only one undamped solution that exists outside the continuum. Its key feature, compared to the single-interface case, is the appearance of a peak whose position in $q_yk_{\rm F}^{-1}$ and height in $\hbar\omega\varepsilon_{\rm F}^{-1}$ depends on both $d$ and $\nu$.

In Fig.~\ref{figBS}(a) we plot two curves, for $d k_{\rm F} = 1$ and $d k_{\rm F} = 2$, which show that the position in $q_y$ and the magnitude of the peak increase with decreasing interface separation. This is due to the fact that, at long wavelengths, a small region, with respect to the plasmon wavelength, will have no effect on it. In this case, the bound mode tracks the continuum region closely and resembles a propagating state due to its poor localisation.

It must be noted, however, that for $|\nu|<2$ and $dk_{\rm F}>2$ the peak loses prominence. Though it remains, since there is also a minuscule peak in the single interface case, it is much suppressed.

Due to the appearance of this peak in the dispersion curve, the group velocity of an interface plasmon would be not only zero at this maximal point but also negative at any point thereafter and very slow for points $q_y\approx q^c_y$. Similar phenomena have been observed in standard metallic thin films.\citep{Fedyanin:2009}

Such a plasmon wavepacket with negative group velocity would then propagate backwards against its initial momentum direction.\cite{Luo:2013} However, due to its shorter wavelength and thus closer proximity to the electron-hole continuum, it would likely decay at a quicker rate than wavepackets of longer wavelengths.

Furthermore, if one were to generate a plasmon wavepacket of a single frequency with constituent waves of momenta (say) $q_1$ and $q_2$ (before and after the peak, respectively) then it would exhibit a beating effect due to the constructive and destructive interference of these constituent waves within the wavepacket.

In Fig.~\ref{figBS}(c) we plot the real part of the potential profile (assuming that quadratic decay terms may be ignored) as a function of the dimensionless $x$ coordinate for an interface separation $dk_{\rm F} = 1$. We show curves for four different values of $q_y$, namely $q_y = \pm0.2 k_{\rm F}$ and $q_y = \pm0.4 k_{\rm F}$. The former (latter) occurring before (after) the turning point of the dispersion curve of Fig.~\ref{figBS}(a).

Interestingly, the potential may be seen to decay across the region in a non-exponential manner reflecting the nature of the plasmon to localise within the region to the interfaces and to then decay outside.

The effect of the interfaces can be seen: the mode transitions from being confined within the whole region $0<x<d$ at small wavevectors, to being completely localised at large $q_y$ to only one of the two interfaces, depending on the combined sign of $\nu q_y$. In the latter case the mode reproduces the short-wavelength limit of the single-interface result, as shown in Fig.~\ref{figBS}(a).
This explains the origin of the turning point in the energy dispersion, which develops because of the transition between these two extremes. 

At all wavelengths the energy of the double-interface mode exceeds that of the single-interface one. This is due to the fact that the two interfaces, due to the opposite jumps in $\delta \sigma_{\rm H}$, have opposite chiralities. As a consequence, if they would be infinitely separated, each of them would host low-energy plasmons propagating in one preferred direction. The latter, being determined by the combined sign of $\nu q_y$, is opposite for the two interfaces. Since the plasmon mode is shared by both of them, one of which has the ``wrong'' chirality, a higher energy is required for it to exist.

This ``wrong chirality" effect may be seen most apparently in the potential plot of Fig.~\ref{figBS}(c). When $\nu q_y>0$ the plasmon localises to the interface at $x=0$. On the other hand, when $\nu q_y<0$ the plasmon not only localises to the other interface but also does so with a reduced amplitude. This effect is as a result of the inclusion of damping and shows the energetically unfavourable nature of the plasmon residing upon the `wrong' interface. So the strongest localisation occurs when $\nu$ and $q_y$ are both negative or both positive.

In Fig.~\ref{figBS}(b) we plot the dimensionless lifetime of the plasmon mode as a function of the wavevector $q_y$. As may be seen, the lifetime possesses a turning point that recedes in $q_y$ as $d$ is increased. However, the lifetime remains roughly constant for all wavevectors and is of order $\tau\gamma\sim1.8$. This minimum corresponds with the point at which the plasmon is shared equally between the two interfaces of opposing chiralities. Hence, its lifetime is negatively affected (albeit minimally) as a result of this energetically unfavourable sharing mechanism.

\section{Propagating States}

Propagating states cannot be found by using the above method. The latter is in fact only applicable for bound states, whose wavevectors satisfy $q_y>q_{\rm p}(\omega)$, and for which the bulk dielectric function $\epsilon(q,\omega)$ [Eq.~(\ref{eq:dielectric_function_bulk})] is non-zero. In the present case, as we will show momentarily, the plasmon wavevector is $q=q_{\rm p}(\omega)$. The bulk dielectric function therefore vanishes, and thus care must be taken in performing the inverse Fourier transform of Eq.~(\ref{eqn:mompot}). Furthermore, since the plasmonic energies considered here are smaller yet than the bound state energies, the approximation that $\omega\gg\Gamma$ will not hold. Thus, a more in depth analysis will be required to include the decay within this section. As such, we now set $\gamma=\Gamma=0$ such that ${\cal J}(x,q_y,\omega)=0$ and so $I(x,q_y,\omega)={\cal I}(x,q_y,\omega)$.

When Eq.~(\ref{eqn:mompot}) by $\epsilon(q,\omega)$, since the latter is zero at $q = q_{\rm p}(\omega)$, we have to introduce terms proportional to the Dirac delta functions $\delta(q\pm q_{\rm p})$ on its right-hand side. The latter indeed vanish when multiplied by $\epsilon(q,\omega)$, returning Eq.~(\ref{eqn:mompot}).

As a result, after this treatment and an inverse Fourier transform, Eq.~(\ref{eqn:mompot}) becomes:
\begin{multline}\label{eqn:proppot}
\phi(x)=-\frac{\delta\sigma_Hq_y}{\omega}\left[I(x,q_y,\omega)\phi(0)-I(x-d,q_y,\omega)\phi(d)\right]
\\
+\frac{C^+}{2\pi}e^{ik_0x}+\frac{C^-}{2\pi}e^{-ik_0x}.
\end{multline}
The last two terms in this equation appear as a result of the added Dirac delta functions. Here $\pm k_0$ are the positions of the poles on the real axis, with $k_0^2(\omega)=q_{\rm p}^2(\omega)-q_y^2$, and $I(x,q_y,\omega)$ is the same integral as defined in Eq.~(\ref{eqn:Iint}). The poles of its integrand are now on the real axis at $\pm k_0$, rather than at $\pm i q_0$ as previously.

In order to determine the scattering coefficients $r$ and $t$ that characterise the magnetic region, we impose that there are no left-moving waves in the region $x>d$, {\it i.e.} terms proportional to $e^{-ik_0x}$ sum to zero. Thus, the potentials in $x<0$ and $x>d$, far from the interfaces, are:
\begin{equation}
\phi(x)=
\begin{cases}
e^{ik_0x}+re^{-ik_0x},\quad & x \to -\infty,\\
te^{ik_0x},\quad & x \to +\infty,
\end{cases}
\end{equation}
where the calculation of $r$ and $t$ is laid out in the appendix. The reflectance and transmittance of the region are then given by $R=|r|^2$ and $T=|t|^2$, respectively, where $R$ and $T$ must satisfy $R+T=1$ by construction.

We express the coefficients $r$ and $t$ in terms of the angle of incidence at $x=0$ using $q_y=q_{\rm p}\sin(\theta)$ and $k_0=q_{\rm p}\cos(\theta)$. They thus read:
\begin{widetext}
\begin{align}
r&=\frac{4\pi i\bigg[\cos(\theta)\left[{\cal I}_B(d,\theta)-{\cal I}_B(0,\theta)\cos\left(\tilde{d}\cos(\theta)\right)\right]
+\left(2\pi-i\tilde{K}^{-1}\cot(\theta)\right)
\sin\left(\tilde{d}\cos(\theta)\right)\bigg]e^{i\tilde{d}\cos(\theta)}}
{\cos^2(\theta)\left[[\tilde{K}\sin(\theta)]^{-2}
+{\cal I}_B^2(d,\theta)-{\cal I}_B^2(0,\theta)\right]
+4i\pi\cos(\theta)\left[{\cal I}_B(0,\theta)-{\cal I}_B(d,\theta)e^{i\tilde{d}\cos(\theta)}\right]+4\pi^2\left[1-e^{2i \tilde{d}\cos(\theta)}\right]},\label{eqn:r}
\\
t&=\frac{\cos^2(\theta)\left[[\tilde{K}\sin(\theta)]^{-2}+{\cal I}_B^2(d,\theta)-{\cal I}_B^2(0,\theta)\right]
+4\pi\cos(\theta){\cal I}_B(d,\theta)\sin[\tilde{d}\cos(\theta)]}
{\cos^2(\theta)\left[[\tilde{K}\sin(\theta)]^{-2}
+{\cal I}_B^2(d,\theta)-{\cal I}_B^2(0,\theta)\right]
+4i\pi\cos(\theta)\left[{\cal I}_B(0,\theta)-{\cal I}_B(d,\theta)e^{i\tilde{d}\cos(\theta)}\right]+4\pi^2\left[1-e^{2i \tilde{d}\cos(\theta)}\right]},\label{eqn:t}
\end{align}
\end{widetext}
where the $\omega$ dependencies are still dropped for brevity. Here we have also introduced $\tilde{K}=\delta\sigma_{\rm H}q_{\rm p}(\omega)/\omega$, $\tilde{d}=q_{\rm p}(\omega)d$ and ${\cal I}_B(x,\theta,\omega)\equiv {\cal I}_B(x,q_y=q_{\rm p}(\omega)\sin(\theta),\omega)$. 

Interestingly, the sign of $\delta\sigma_{\rm H}$ has no effect on the reflectance, $R=|r|^2$, and transmittance, $T=|t|^2$, of the region. This is as a result of its appearance in $r$ and $t$ as part of either a squared term or an imaginary term.

As a final note, we have presented $r\equiv r_{\rm L}$ and $t\equiv t_{\rm L}$, {\it i.e.} the scattering coefficients from left to right. However, due to the mirror symmetry of the magnetic region, it follows that $r_{\rm R}=r$ and $t_{\rm R}=t$.

\begin{figure}
\begin{center}
\begin{overpic}[width=1.0\columnwidth]{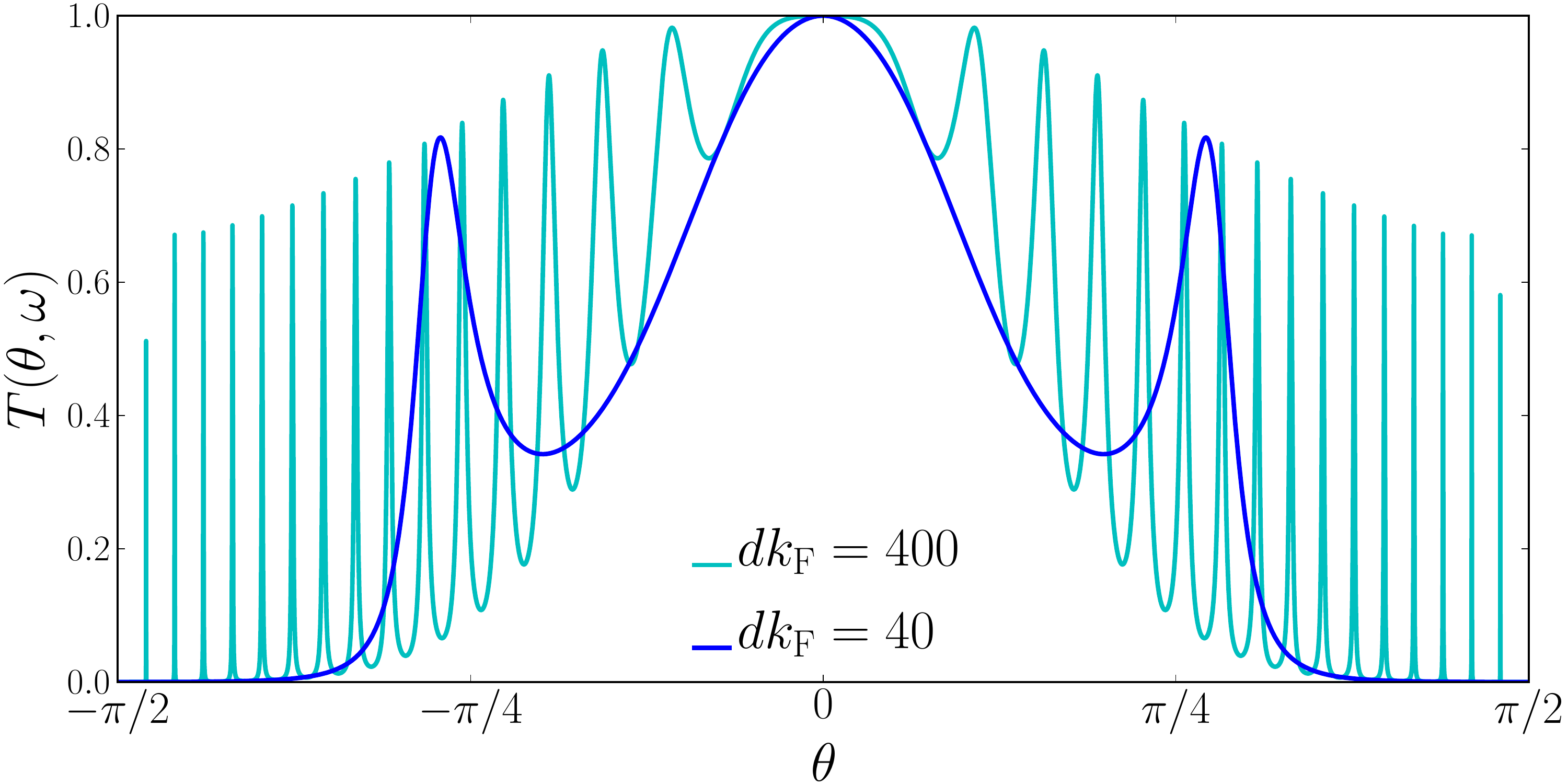}\put(8,45.5){(a)}
\end{overpic}
\begin{overpic}[width=1.0\columnwidth]{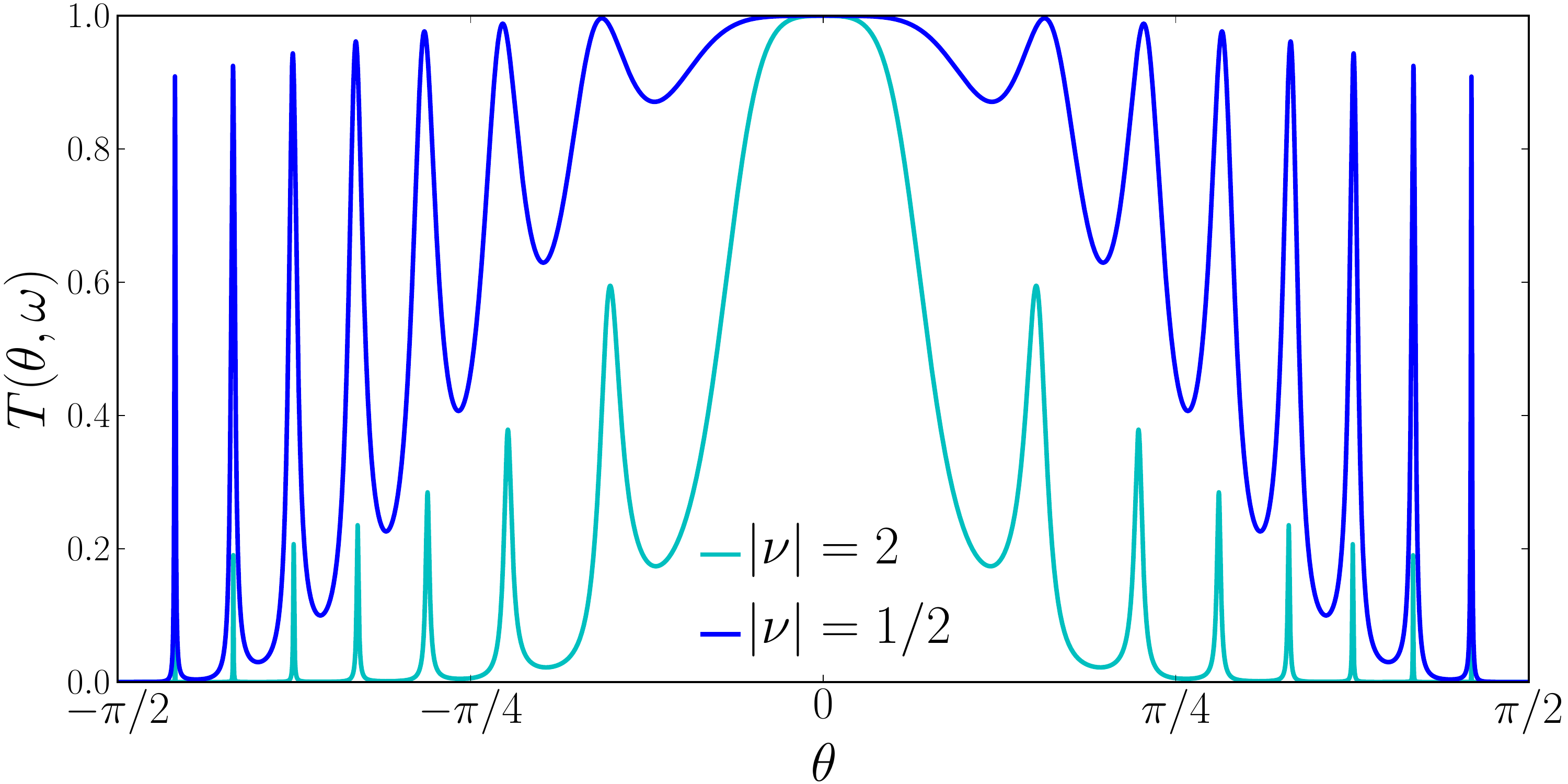}\put(8,45.5){(b)}
\end{overpic}
\end{center}%
\caption{(Colour on-line)
Panel (a): The number of transmittance peaks at fixed plasmon frequency increases with decreasing dimensionless interface separation $d k_{\rm F}$. In this plot, the plasmon frequency is $\hbar\omega=\varepsilon_{\rm F}/2$ and $|\nu|=1$. Importantly, the peak strengths do not decay heavily as the angle is increased for small $\nu$, this is shown further in the $|\nu|=1/2$ case in panel (b).
Panel (b): The maximum transmittance of the side peaks is seen to decrease whilst they become much sharper with increasing $|\nu|$. Here, $\hbar\omega=\varepsilon_{\rm F}/2$ and $dk_{\rm F}=200$. This strongly improves the quality factors of all transmission peaks but suppresses the intensity of any side peaks, {\it i.e.} a large $\nu$ causes the film to become a strong mirror except in a narrow region around $\theta\approx0$ and a damped selector for certain angles besides. On the other hand, a small $\nu$ causes the film to become transparent as the quality factors of side peaks decreases.}
\label{figPS}
\end{figure}

In Fig.~\ref{figPS} we plot the transmittance of propagating modes as a function of the plasmon angle of incidence $\theta$ from the normal to the interface. In panel (a) we show two curves for two distinct values of the dimensionless interface separation $d k_{\rm F}=40$ and $dk_{\rm F}=400$ with the same $\nu$ parameter $|\nu|=1$.

The interface separation has a dramatic effect: small-width regions exhibit selective angle-dependent transmission, however with rather poor quality factor. On the other hand, for large-width regions, many sharp side transmission peaks are seen to appear. The central peak remains broad in both cases. Furthermore, we find that if $dk_{\rm F}<40$, the peaks disappear. This is because the plasmon will again not see the region and will instead propagate through it unaffected. Note that the spectrum shows the typical ``Airy-disk'' characteristic of Fabry-P{\'e}rot resonance, wherein the linewidth is directly related to the region width.\cite{Soto:2016,Cox:1992} 

In panel (b) we plot instead the transmittance as a function of the incident angle but for a fixed $dk_{\rm F}=200$ and two values of the parameter $|\nu|$: $|\nu|=1/2$ and $|\nu|=2$. In this case it can be seen that the quality factor (sharpness) of all peaks is increased. However, the transmittance of the side peaks is suppressed as a result of the increasing $\nu$ parameter.

Thus there is a trade-off. To have transmission peaks with high quality factor, the $\nu$ parameter must be large yet this increase diminishes the strength of said peaks. The same goes by changing the width $d$. The quality factors of the peaks increase as $d$ increases. However, the peaks become closer to each other, and hence more and more difficult to resolve.

\section{Summary and conclusions}
To summarise, we developed a semi-classical description of plasmonic excitations in the presence of a frequency independent step-wise-varying off-diagonal Hall conductivity. 

We found that a plasmon can propagate confined between the interfaces. For a given energy, said plasmon has a larger wavevector than the bulk ones, and therefore can be excited separately.
Its energy dispersion shows a turning point at which the plasmon has zero group velocity. The mode is bound to one of the interfaces depending on the combined sign of the momentum along the interface and the ``filling factor'' $\nu$ that parametrizes the differences in Hall conductivities between the regions, in units of $e^2/h$. The bound plasmon also shows a typical localisation length which is inversely proportional to this momentum. 
By studying the scattering process for an incident plasmon through the region, we calculate the reflection and transmission coefficients. The number of side transmission peaks depends heavily on the interface separation, whilst their intensity and sharpness decrease with the parameter $\nu$.

As can be seen in figure \ref{figBS}, the interface state localises very strongly to the region as a whole with preference for either of the interfaces, depending on the sign of the jump is Hall conductance, as the wavevector increases. As such, the thin film geometry could find application as a plasmonic waveguide.

\begin{figure}
\centering
\includegraphics[width=\linewidth]{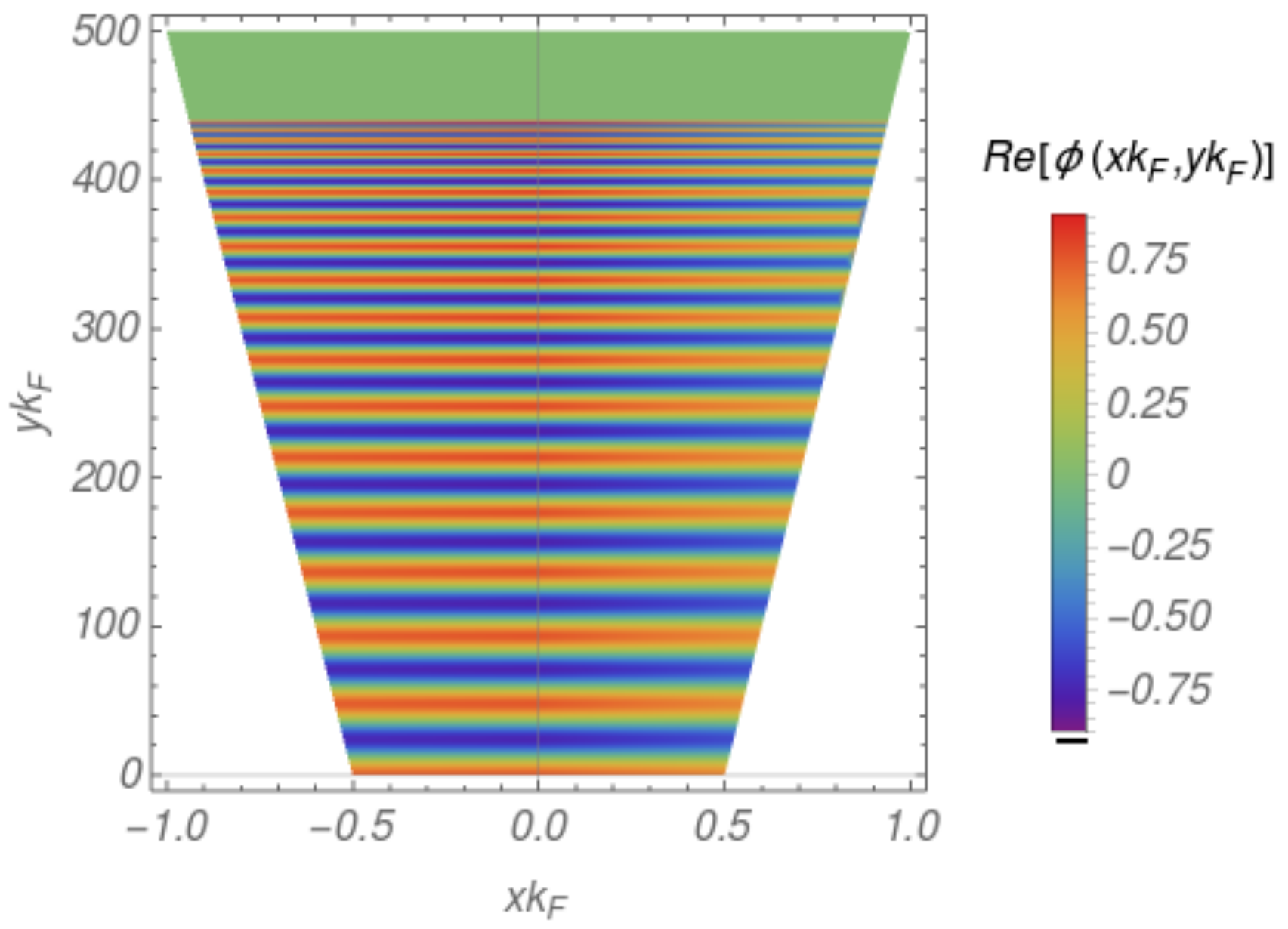}
\caption{A contour plot of the potential profile of a bound plasmon within an adiabatically widening thin-film. Here, for consistency, the parameters are identical to those of Fig.~\ref{figBS}: $\nu=2$, $\omega=0.48\varepsilon_{\rm F}$ with $q_y$ as the smaller positive solution to Eq.~(\ref{eqn:dispersion}) with this chosen frequency. The width of the region is varied from $dk_{\rm F}=1$ to $dk_{\rm F}=2$ over a suitably large range of $y$: $0\leq yk_{\rm F}\leq500$. The localisation of the plasmon to the left-hand side of the thin-film may be seen (albeit faintly) in addition to the `invisible' barrier to its $\hat{\bm{y}}$-directional propagation whereat the plasmon ceases to exist and so `reflects' back.}
\label{figSP}
\end{figure}

The fact that the bound state dispersion curve exhibits a maximum, at which point the group velocity vanishes, may be exploited to confine interface plasmons within a finite region without the need of a solid barrier. We recall indeed that the wavevector at which the plasmon dispersion peaks, as well as the peak energy, depend on the geometrical parameters of the structure. In particular, the peak lowers in frequency with increasing interface separation. Therefore, one could imagine to shape the region in such a way that a wavepacket with a fixed frequency will eventually stop propagating and bounce back when the group velocity vanishes. This is achieved by adiabatically increasing the interface separation away from the point where the wavepacket is created in such a way that its dispersion evolves adiabatically while it propagates. If the thin film widens in both directions, then the wavepacket would be confined within the region and would thus become a confined standing wave. Such a plasmon may be seen in Fig.~\ref{figSP}.

For \textit{propagating} modes, the fact that the side transmission peaks may be modulated in number, intensity and quality factor through the variation of $d$, $\nu$ and $\omega$ could be used to generate monochromatic plasmons. By constructing a resonator with a given width $d$ and $\nu$, a plasmon with a certain frequency may be made to pass through alone by sending it at a specific angle $\theta$. Thus, incident plasmons of specific frequencies may be selected for by detecting them at an angle after the magnetic region. Furthermore, the opacity of the region to plasmons of certain energies depending on $d$ and $\nu$ could be used to confine plasmons of such energies between two regions. However, if a plasmon were to lose energy during its propagation, {\it i.e.} through any form of decay, then the regions would appear transparent to the plasmon at certain energies thus hampering the confinement quality.

Finally, we wish to comment on the feasibility of our set-up. Candidates for the realisation of these phenomena are metallic Dirac-like 2D surface states (e.g., those at the surface of a 3D TI, accounting for the proper dielectric environment) as mentioned in the introduction. Such systems exhibit typical surface electronic number densities of $n\sim\SI{e12}{\centi\metre^{-2}}$ and a Fermi energy of $\varepsilon_{\rm F}=\hbar v_{\rm F}k_{\rm F}$. Thus, the Fermi momentum (in units of $\sqrt{N_F}$) for the system is given by $k_{\rm F}=\sqrt{4\pi n}\sim\SI{3.5e6}{\centi\metre^{-1}}=\SI{0.35}{\nano\metre^{-1}}$. Moreover, their typical Fermi velocity is $v_{\rm F}\sim\SI{e8}{\centi\metre\second^{-1}}$ and so the Fermi energy is $\varepsilon_{\rm F}\sim\SI{0.23}{\electronvolt}$. (Note that: taking $N_F=4$, as in graphene, would halve both the Fermi momentum and the Fermi energy.) Finally, taking  a typical experimental electron scattering time of $\tau_{\rm sc}=\SI{50}{\femto\second}$,\cite{Yin:2017} we may see that the lifetime of the bound plasmon is $\tau\sim1.8\tau_{\rm sc}$ and so $\tau\approx\SI{100}{\femto\second}=\SI{0.1}{\pico\second}$. This result is at least two orders of magnitude smaller than the results of Ref.~\onlinecite{Hasdeo:2017}.
However, these numbers are merely used as ball-park figures in order to convey a general sense of the scales involved in the system. These quantities may all be modulated at will given suitable materials or to suit certain experimental conditions.

Considering these ball-park figures, observation of zero/negative group velocity for the bound interface states would require separations of $d\lesssim\SI{1}{\nano\metre}$ and a change in AQH conductivity of $|\nu|\gtrsim2$. For $dk_{\rm F}\gg1$, the second interface becomes irrelevant so the mode localises to the single dominant interface. Whereas, for $d<\SI{1}{\nano\metre}$, the semi-classical method breaks down as effects due to the underlying crystal lattice begin to dominate and the position of the zero group velocity turning point tends towards the electron-hole continuum that begins at $q_y\sim k_{\rm F}$. Such a large step-wise change in the Hall conductivity is also unlikely to be able to implemented experimentally.

The use of the studied heterostructure as a plasmonic waveguide has much better chances. We find that plasmons can be bound within magnetic strips satisfying $d\gtrsim20k_{\rm F} \sim \SI{7}{\nano\metre}$ and $|\nu|<1$. In this case, the bound plasmon would localise to either of the interfaces, depending on the sign of $\nu q_y$, rather than inside the region, as explained above. For, say positive $\nu$, a plasmon with $q_y>0$ moving `up' the region would localise to the interface at $x=0$ whilst a plasmon with $q_y<0$ moving `down' would localise to the other interface at $x=d$. Yet, the rather small lifetime might render it difficult to utilise. Nevertheless observation ought not to be impossible.

Peaked transmission spectra require rather large separations in the range $\SI{40}{\nano\metre}<d\lesssim\SI{400}{\nano\metre}$, well within the studied semi-classical regime. Heterostructures working as plasmon filters could therefore be well realised experimentally, and their theoretical description does not require the consideration of quantum effects.  The upper limit of $\SI{400}{\nano\metre}$ is not a strong one since a larger region would simply see an increase in the number of peaks in the transmission spectrum. Admittedly, when the number of peaks becomes too large they blur together and cease to be resolvable, thus making the region practically transparent for all angles. Conversely, the lower limit of $\SI{40}{\nano\metre}$ is a stringent one: below that, peaks do not occur. 
Frequency selection of plasmons could well be seen within experimental conditions. In fact, typical plasmonic energies of metallic surfaces are of order $\hbar\omega\sim \varepsilon_{\rm F}/2 \sim \SI{0.1}{\electronvolt}$,\cite{Woessner:2014} and therefore observable under typical experimental conditions.

\section{Acknowledgements}

We would like to thank the referees for their pertinent criticisms and valuable suggestions without which this work would be of a considerably lower standard.

T.B.S. acknowledges the support of the EPSRC through a PhD studentship grant. A.P. and T.B.S. acknowledge support from the Royal Society International Exchange grant IES\textbackslash R3\textbackslash 170252.

\bibliographystyle{apsrev4-1}
\bibliography{bibliography}

\onecolumngrid
\appendix
\section{Calculation of the scattering coefficients $r$ and $t$}
The boundary conditions that specify $C^+$ and the ratio of $\phi(0)$ and $\phi(d)$ are generated by the imposition of continuity of the potential at the interfaces, whilst $C^-$ is specified by the imposition that there are only right-moving waves in $x>d$. By evaluating Eq.~(\ref{eqn:proppot}) at $x=0$ and $x=d$ we find:
\begin{align}
\frac{1}{2\pi}\left(C^++C^-\right)&=\left[1+KI_B(0)\right]\phi(0)
-K\left[2\pi\frac{q_{\rm p}}{k_0}\sin(k_0d)+I_B(d)\right]\phi(d),\label{eqn:Ccond1}
\\
\frac{1}{2\pi}\left(C^+e^{ik_0d}+C^-e^{-ik_0d}\right)&=\left[1-KI_B(0)\right]\phi(d)
+K\left[2\pi\frac{q_{\rm p}}{k_0}\sin(k_0d)+I_B(d)\right]\phi(0),\label{eqn:Ccond2}
\end{align}
where any dependence on $q_y$ and $\omega$ in $q_{\rm p}(\omega)$, $k_0(\omega)$ and $I_B(x,q_y,\omega)$ has been dropped for brevity and $K=\delta\sigma_{\rm H}q_y/\omega$.

To find the reflectance and transmittance through the central region from $x<0$ to $x>d$, we impose that there are no left moving waves ($\propto e^{-ik_0x}$) in the region $x>d$, \textit{i.e.} (from Eq.~(\ref{eqn:proppot})):
\begin{equation}\label{eqn:Cminus}
\frac{C^-}{2\pi}=i\pi K\frac{q_{\rm p}}{k_0}\left[\phi(0)-\phi(d)e^{ik_0d}\right].
\end{equation}
Then, far from the interfaces, the potentials have the following form:
\begin{equation}
\phi(x)=
\begin{cases}
Ae^{ik_0x}+Be^{-ik_0x},\quad & x \to -\infty,\\
Ce^{ik_0x},\quad & x \to +\infty,
\end{cases}
\end{equation}
such that the reflection and transmission coefficients are simply $r=B/A$ and $t=C/A$. $A$, $B$ and $C$ may then be found by solving Eqs.~\eqref{eqn:Ccond1} and \eqref{eqn:Ccond2} for $C^+$ and $\phi(0)/\phi(d)$, with $C^-$ given by Eq.~\eqref{eqn:Cminus}, and then plugging the results back into Eq.~\eqref{eqn:proppot}.

Firstly, solving for $C^+$ by substituting Eq.~(\ref{eqn:Cminus}) into Eq.~(\ref{eqn:Ccond1}) yields:
\begin{equation} \label{eqn:Cplus}
\frac{C^+}{2\pi}=\left[1+KI_B(0)-i\pi K\frac{q_{\rm p}}{k_0}\right]\phi(0)+K\left[i\pi\frac{q_{\rm p}}{k_0}\left(2e^{ik_0d}-e^{-ik_0d}\right)-I_B(d)\right]\phi(d).
\end{equation}
Then, secondly, the ratio $\phi(0)/\phi(d)$ may be determined by using Eqs.~(\ref{eqn:Ccond2},\ref{eqn:Cminus},\ref{eqn:Cplus}) together as:
\begin{equation}\label{eqn:phirat}
\frac{\phi(0)}{\phi(d)}=\frac{k_0\left(1+K\left[I_B(d)e^{ik_0d}-I_B(0)\right]\right)-2i\pi Kq_{\rm p}\left(e^{2ik_0d}-1\right)}{k_0\left(\left[1+KI_B(0)\right]e^{ik_0d}-KI_B(d)\right)}.
\end{equation}
Thus, $A$, $B$ and $C$ may be found now as the appropriate coefficients of $e^{\pm ik_0x}$ as in Eq.~(\ref{eqn:proppot}). Explicitly, we have:
\begin{align}
A&=\frac{C^+}{2\pi}-i\pi K\frac{q_{\rm p}}{k_0}\left[\phi(0)-\phi(d)e^{-ik_0d}\right],
\\
B&=\frac{C^-}{2\pi}+i\pi K\frac{q_{\rm p}}{k_0}\left[\phi(0)-\phi(d)e^{ik_0d}\right],
\\
C&=\frac{C^+}{2\pi}+i\pi K\frac{q_{\rm p}}{k_0}\left[\phi(0)-\phi(d)e^{-ik_0d}\right],
\end{align}
and thus, through the use of Eqs.~(\ref{eqn:Cminus},\ref{eqn:Cplus},\ref{eqn:phirat}) and after some lengthy algebra, we arrive at:
\begin{align}
A&=\left[\frac{k_0^2\left(1+K^2\left[I_B^2(d)-I_B^2(0)\right]\right)+4i\pi  K^2q_{\rm p}k_0\left[I_B(0)-e^{ik_0d}I_B(d)\right]+4\pi^2K^2q_{\rm p}^2\left(1-e^{2ik_0d}\right)}{k_0^2\left(e^{ik_0d}+K\left[I_B(0)e^{ik_0d}-I_B(d)\right]\right)}\right]\phi(d),
\\
B&=\left[\frac{4i\pi Kq_{\rm p}\left[Kk_0\left[I_B(d)-I_B(0)\cos(k_0d)\right]+\left(2\pi Kq_{\rm p}-ik_0\right)\sin(k_0d)\right]e^{ik_0d}}{k_0^2\left(e^{ik_0d}+K\left[I_B(0)e^{ik_0d}-I_B(d)\right]\right)}\right]\phi(d),
\\
C&=\left[\frac{k_0\left(1+K^2\left[I_B^2(d)-I_B^2(0)\right]+4\pi K^2q_{\rm p}I_B(d)\sin(k_0d)\right)}{k_0^2\left(e^{ik_0d}+K\left[I_B(0)e^{ik_0d}-I_B(d)\right]\right)}\right]\phi(d),
\end{align}
from which  we find the scattering coefficients as $r=B/A$ and $t=C/A$. Finally, taking $q_y=q_{\rm p}(\omega)\sin(\theta)$ and $k_0=q_{\rm p}(\omega)\cos(\theta)$ along with simple rearrangement yields the forms of $r$ and $t$ as quoted in Eqs.~(\ref{eqn:r},\ref{eqn:t}).

The same analysis may be applied to the reverse case where the scattering occurs from right to left. The result may be seen to be identical in such a case: $r_{\rm L}=r_{\rm R}=r$ and $t_{\rm L}=t_{\rm R}=t$, due to the mirror symmetry of the region in the line $x=d/2$.

\end{document}